\documentclass{emulateapj}
\newcommand{\beq}{\begin{equation}}
\newcommand{\beqa}{\begin{eqnarray}}
\newcommand{\eeq}{\end{equation}}
\newcommand{\eeqa}{\end{eqnarray}}
\newcommand{\etal}{{\it et al. }}
\newcommand{\bfx}{\mathbf{x}}
\newcommand{\bfk}{\mathbf{k}}

\newcommand{\mobs}{\mathrm{obs}}

\newcommand{\mi}{\mathrm{i}}

\newcommand{\ms}{\mathrm{s}}
\newcommand{\me}{\mathrm{e}}

\newcommand{\mc}{\mathrm{c}}

\slugcomment{}

\shorttitle{Amplitude and Phase Fluctuations for GWs}
\shortauthors{R. Takahashi}

\begin{document}


\title{Amplitude and Phase Fluctuations for Gravitational Waves
 Propagating through Inhomogeneous Mass Distribution in the Universe}


\author{Ryuichi Takahashi}
\affil{Division of Theoretical Astronomy, National
               Astronomical Observatory of Japan, Mitaka, 
               Tokyo 181-8588, Japan}
\email{takahasi@th.nao.ac.jp}


\begin{abstract}

When a gravitational wave (GW) from a distant source propagates through
 the universe, its amplitude and phase change due to gravitational
 lensing by the inhomogeneous mass distribution.
We derive the amplitude and phase fluctuations, and calculate
 these variances in the limit of a weak gravitational field of
 density perturbation.
If the scale of the perturbation is smaller than the Fresnel scale
 $\sim 100 \mbox{pc} (f/\mbox{mHz})^{-1/2}$ ($f$ is the GW frequency),
 the GW is not magnified due to the diffraction effect.
The rms amplitude fluctuation is $1-10 \%$ for $f > 10^{-10}$ Hz,
 but it is reduced less than $5 \%$ for a very low frequency of
 $f < 10^{-12}$ Hz.
The rms  phase fluctuation in the chirp signal is $\sim 10^{-3}$ radian at
 LISA frequency band ($10^{-5} - 10^{-1}$ Hz).
Measurements of these fluctuations will provide information about
 the matter power spectrum on the Fresnel scale
 $\sim 100$ pc.

\end{abstract}


\section{Introduction}

The inspiral and merger of SMBHs (Super Massive Black Holes : $10^4-10^7 
 M_\odot$) is one of the most
 promising candidates for LISA (Laser Interferometer Space Antenna),
 which will be launched around $2014$. 
This detector is sensitive in a frequency range of $10^{-5}-10^{-1}$ Hz
 and can measure SMBH mergers at cosmological distances with a
 high signal-to-noise ratio. 
The binary SMBH systems have recently been called ``cosmological
 standard sirens'' (Holz \& Hughes 2005).
This is because the distance to the source $r_\ms$ can be directly measured
 by using the relation $A \propto \dot{f}/(f^3 r_\ms)$, where
 $f$  is the GW frequency, $\dot{f}$ is its time derivative and $A$ is
 the amplitude at the detector (Schutz 1986).
The distance can be determined with
 less than $1 \%$ accuracy if the direction is determined by
 identifying its electromagnetic counterpart (e.g. Cutler 1998; Seto 2002;
 Hughes 2002; Vecchio 2004).

But in practice, the distance cannot be determined with such high
 accuracy because of the gravitational lensing caused by
 inhomogeneous mass distribution in the Universe.
Recently, Holz \& Hughes (2005) and Kocsis et. al. (2005) discussed the
 effects of lensing magnification (or demagnification) on determining
 the distance to SMBH mergers.
They concluded that lensing errors are $5-10 \%$, which is greater than
 the intrinsic distance error.

In weak gravitational lensing, the magnification is
 $\mu \simeq 1 + 2 \kappa$, where $\kappa$ is the convergence.
The rms convergence fluctuation was derived by
 Blandford et al. (1991), Miralda-Escude (1991) and Kaiser (1992) 
 on the basis of linear perturbation theory.
In this previous work, the geometrical optics was assumed.

But for the lensing of GWs, since the wavelength is much longer than that
 of light, the geometrical optics is not valid in some cases. 
Recently, suggested by Macquart (2004) and Takahashi, Suyama \&
 Michikoshi (2005), if the scale of the density perturbation
 $k^{-1}$ is smaller than the Fresnel scale $r_F \sim 
 (\lambda r_\ms)^{1/2}$ ($\lambda$ is the wavelength),
 the wave effects become important.
This condition is rewritten as
 $k^{-1} < 100 \mbox{pc} ( {f}/{\mbox{mHz}} )^{-1/2}
 ( {r_\ms}/{10 \mbox{Gpc}} )^{1/2}$.
In such a case, the incident wave does not experience
 perturbation and its amplitude is not magnified.

In this paper, we consider a situation in which GWs propagate through
 the density perturbation of CDM (cold dark matter) and baryon.
In the wave optics, the lensing affects not only the amplitude but also
 the phase, hence we discuss the lensing effects on both of them.
We use units of $c=G=1$.

\section{Gravitationally Lensed Waveform}

\begin{figure}
\epsscale{0.8}
\plotone{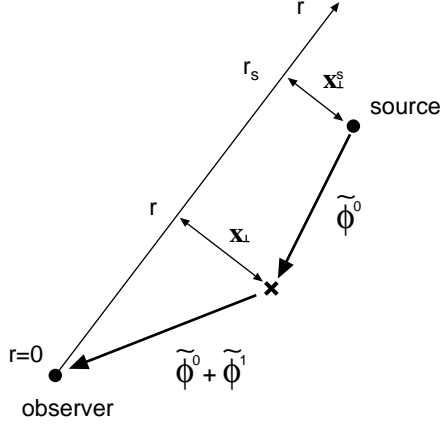}
\caption{Gravitational lens geometry for the observer and the source.
$r$ is the comoving distance from the observer, $r(z)=\int_0^z dz^\prime
 /H(z^\prime)$, and $\bfx_\perp$ is a two-dimensional
 vector perpendicular to the line-of-sight.
The source position is $\bfx_\ms=(r_s,\bfx_\perp^\ms)$ with
 $|\bfx_\perp^\ms| \ll r_\ms$.
$\tilde{\phi}^0$ is the incident wave, and $\tilde{\phi}^0+\tilde{\phi}^1$
 is the lensed wave.}
\label{config}
\vspace{0.8cm}
\end{figure}

The background metric is the Friedmann-Robertson-Walker (FRW) model
 with a gravitational potential of $U (\ll 1)$.
The perturbed FRW metric $\tilde{g}^B_{\mu \nu}$ for a flat universe
 is given by
\beqa
  d\tilde{s}^2 &=& \tilde{g}^B_{\mu \nu} dx^\mu dx^\nu
 = a^2(\eta) g^B_{\mu \nu} dx^\mu dx^\nu  \nonumber \\
 &=& a^2(\eta) \left[ -\left( 1+2 U \right) d\eta^2 + \left( 1-2 U \right)
 d\bfx^2 \right],
\label{metric}
\eeqa
 where $\eta$ is the conformal time, the scale factor is normalized
 such that $a=1$ at present, and $g^B_{\mu \nu}$ is the perturbed
 Minkowski metric.
The line element is $d\bfx^2=dr^2+d\bfx_\perp^2$, where
 $r$ is a radial coordinate (a comoving distance) from the observer,
 $r(z)=\int_0^z dz^\prime/H(z^\prime)$, 
 while $\bfx_\perp$ is a two-dimensional
 vector perpendicular to the line-of-sight.
We show the lens geometry for the observer and the source in
 Fig.\ref{config}.
The observer is in the origin of the coordinate axes, while
the source position is $\bfx_\ms=(r_s,\bfx_\perp^\ms)$ with
 $|\bfx_\perp^\ms| \ll r_\ms$.

Since the propagation equation of GW is (1) conformally
 invariant if the wavelength is much smaller than the Hubble radius
 (see Appendix) and (2) the same as the scalar field wave equation
 (Peters 1974),
 we use the scalar field $\phi$ propagating under the Minkowski
 background $g^B_{\mu \nu}$.
The basic equation is
 $\partial_\mu \left( \sqrt{-g^B} g_B^{\mu \nu} \partial_\nu \phi \right)
 =0$ or
\beq
\left( \nabla^2+\omega^2 \right) \tilde{\phi}
 = 4 \omega^2 U \tilde{\phi},
\label{wv}
\eeq
 where $\omega$ is the frequency at the
 observer\footnote{$\omega$ is the same as the comoving
 frequency $\omega_{\mbox{c}}$, since $\omega_{\mbox{c}}=a \omega$ and
 $a=1$ at present.}
 and $\tilde{\phi}(\omega,\bfx)$ is the Fourier transform of
 $\phi(\eta,\bfx)$.

We take $\tilde{\phi}^0$ as the the incident wave emitted from the source,
 which is the solution of Eq.(\ref{wv}) in the unlensed case $U=0$.
We use the spherical wave as $\tilde{\phi}^0$ : 
 $\tilde{\phi}^0(\omega,\bfx) \propto
 \me^{\mi \omega |\bfx-\bfx_\ms|}/|\bfx-\bfx_\ms|$.
Including the effect of $U$ on the first order (Born approximation),
 the gravitationally lensed wave at the observer is given by \cite{tsm05}
\beq
  \tilde{\phi}^L_\mobs(\omega)=\tilde{\phi}^0_\mobs(\omega)
 +\tilde{\phi}^1_\mobs(\omega),
\label{phi}
\eeq
with
\beq
  \tilde{\phi}^1_\mobs(\omega)=-\frac{\omega^2}{\pi} \int d^3x
 \frac{\me^{\mi \omega \left| \bfx \right|}}
 {\left| \bfx \right|} U(\bfx)
 \tilde{\phi}^0(\omega, \bfx).
\label{sw}
\eeq
where $\tilde{\phi}^1$ represents the effect of lensing ($|\tilde{\phi}^1|
 \ll |\tilde{\phi}^0|$).
The incident wave $\tilde{\phi}^0$
 is gravitationally lensed at $\bfx=(r,\bfx_\perp)$ and
 changed into the lensed wave $\tilde{\phi}^0+\tilde{\phi}^1$
 (see Fig.\ref{config}).

Let us define $K$ and $S$ as (Ishimaru 1978, Ch.17),
\beq
  K(\omega) = \mathrm{Re} \left[ \frac{\tilde{\phi}^1_\mobs(\omega)}
 {\tilde{\phi}^0_\mobs(\omega)} \right];
 ~S(\omega) = \mathrm{Im} \left[ \frac{\tilde{\phi}^1_\mobs(\omega)}
 {\tilde{\phi}^0_\mobs(\omega)} \right].
\label{as}
\eeq
Then we have
\beq
 \tilde{\phi}^L_\mobs(\omega)=\left[1+K(\omega)\right] \tilde{\phi}^0_\mobs
 (\omega) \me^{\mi S(\omega)},
\eeq
 from Eq.(\ref{phi}).
Hence $K$ means the magnification ($K>0$) or demagnification ($K<0$) of
 the wave amplitude, while $S$ means the phase shift due to the lensing.
Hereafter, we call $K$ {\it amplitude fluctuation} and $S$ 
 {\it phase fluctuation},
 respectively.

Using the Fourier transform of the potential, $\tilde{U}(r,\bfk)
 = \int d^3xU(\bfx) \me^{\mi \bfk \cdot \bfx}$,
 with $|\bfx_\perp| \ll r$ and $|\bfx_\perp^\ms| \ll r_\ms$,
 the result in Eq.(\ref{sw}) is reduced to
\beqa
  \frac{\tilde{\phi}^1_\mobs(\omega)}{\tilde{\phi}^0_\mobs(\omega)} =
 -2 \, \mi \, \omega
 \int_0^{r_\ms} dr \int \frac{d^3 k}{(2 \pi)^3} ~\tilde{U}(r, \bfk)
 \nonumber \\ \times \exp \left[ -\mi k_r r
  -\mi \frac{r}{r_\ms} \bfk_\perp \!\! \cdot \! \bfx_\perp^\ms
 -\mi \frac{r \left( r_\ms-r \right)}{2 \omega r_\ms}
 \left| \bfk_\perp \right|^2 \right],
\label{sw2}
\eeqa
where $k_r$ and $\bfk_\perp$ are the radial and perpendicular
 components of $\bfk$.
In particular for high frequency limit $\omega \to \infty$, $K$ and $S$
 are rewritten as
\beqa
 K(\omega) &\to& \kappa = \int_0^{r_\ms} dr
 \frac{r \left( r_\ms-r \right)}{r_\ms}
 \nabla^2_\perp U(r, \frac{r}{r_\ms} \bfx_\perp^\ms), \nonumber \\
 S(\omega) &\to& \omega t_d = -2 \omega \int_0^{r_\ms} dr
 U(r,\frac{r}{r_\ms} \bfx_\perp^\ms).
\label{asgeo}
\eeqa
Here, $\kappa$ is the convergence field along the line-of-sight to 
 the source and $t_d$ is the gravitational time-delay.  
The above results are consistent with that in weak gravitational
 lensing \cite{bs01}.

\section{Amplitude Fluctuation}

\subsection{Variance in the Amplitude Fluctuation}

In this section, we derive the variance in the amplitude fluctuation $K$.
The gravitational potential satisfies Poisson's equation (Peebles 1980)
\beq
 \tilde{U}(r,\bfk) = - \frac{3 H_0^2 \Omega_0}{2 a(r)}
 k^{-2} \tilde{\delta}(r,\bfk),
\eeq
where $\tilde{\delta}$ is the density perturbation.
The fluctuation of $\tilde{\delta}$ is characterized by the
 power spectrum : $\langle
 \tilde{\delta}(r,\bfk) \tilde{\delta}^*(r^\prime,\bfk^\prime) \rangle
 = (2\pi)^4 P_\delta(r,k) \delta^3(\bfk-\bfk^\prime) \delta(r-r^\prime)$.
Then, the correlation in the potentials $\tilde{U}(r,\bfk)$ and
 $\tilde{U}(r^\prime,\bfk^\prime)$ is
\beqa
 \langle \tilde{U}(r,\bfk)\tilde{U}(r^\prime,\bfk^\prime) \rangle =
 \left[ \frac{3 H_0^2 \Omega_0}{2 a(r)} \right]^2 k^{-4} \nonumber \\
  \times \left( 2 \pi \right)^4 P_\delta(r,k)
 \delta^3(\bfk-\bfk^\prime) \delta(r-r^\prime).
\label{pk}
\eeqa
To calculate $P_\delta$, we use the linear power spectrum \cite{eh99}
 with the nonlinear correction of Peacock \& Dodds (1996).
We adopt a COBE-normalized\footnote{It corresponds to $\sigma_8=0.8$
 which is the present amplitude of
 the mass fluctuation at $8 h^{-1}$ Mpc.},
 scale invariant ($n=1$) power spectrum in
 a flat $\Lambda$CDM cosmology with $\Omega_b=0.04$, $\Omega_0=0.3$,
 $\Omega_\Lambda=0.7$, and
 $H_0=70~\mbox{km s}^{-1} \mbox{Mpc}^{-1}$.

\begin{figure}
\plotone{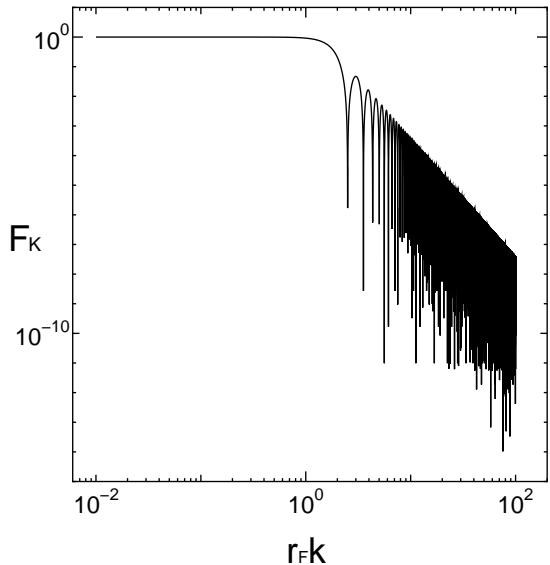}
\caption{Filter function $F_K$ as a function of $r_F k$.}
\label{win}
\end{figure}

The variance in the amplitude fluctuation is given from Eqs. (\ref{as}),
 (\ref{sw2}) and (\ref{pk}) as
\beqa
  \Delta_K^2(\omega) \equiv \langle K^2(\omega) \rangle =
 \left( \frac{3 H_0^2 \Omega_0}
 {4 \pi} \right)^2 \int_0^{r_\ms} dr \frac{1}{a^2(r)}
 \nonumber \\
  \times \frac{r^2 \left( r_\ms-r \right)^2}{r_\ms^2}
 \int d^2 k_\perp
 P_\delta (r,k_\perp) F_K(\omega,r,k_\perp)
\label{dela}
\eeqa
where the filter function $F_K$ is defined as
\beq
  F_K= \left[ \frac{\sin \left( r_F^2 k_\perp^2 /2 \right)}
 {r_F^2 k_\perp^2 /2} \right]^2 ; ~r_F^2=\frac{r \left( r_\ms-r\right)}
 {\omega r_\ms}.
\label{fres}
\eeq
Here, $r_F$ is called the Fresnel scale \cite{m04}.
This is roughly given by
\beq
  r_F \simeq 120 \mbox{pc}  \left( \frac{f}{\mbox{mHz}} \right)^{-1/2}
  \left[ \frac{r \left( r_\ms-r \right)/r_\ms}{10 \mbox{Gpc}}
 \right]^{1/2},
\eeq
where $f=\omega/2\pi$.
We show the filter function $F_K$
 as a function of $r_F k_\perp$ in Fig.\ref{win}.
From this figure, $F_K=1$ if the scale of the density
 perturbation $\sim k^{-1}_\perp$ is larger than the Fresnel scale $r_F$,
 while $F_K$ decreases rapidly in proportion to $(r_F k_\perp)^{-4}$
 if $k^{-1}_\perp \ll r_F$.
Hence, the amplitude fluctuation is affected by a density perturbation of
 scale larger than the Fresnel scale.
In the geometrical optics limit $F_K \to 1$,
 the result (\ref{dela}) is the same as $\langle \kappa^2 \rangle$. 

The reason why a critical scale is $r_F$ can be explained as follows : 
For the lensing by a compact object with a mass $M$, if the wavelength
 is larger than the Schwarzschild radius, $\lambda > M$, the diffraction
 effect becomes important (see Takahashi \& Nakamura 2003, and references
 therein).
Inserting this condition to
 the Einstein radius $r_E=[4 M (r_\ms-r)/r_\ms]^{1/2}$,
% which is identified with the scale of the perturbation 
 we have the Fresnel scale in Eq.(\ref{fres}).

%The reason why a critical scale is $r_F$ can be explained as follows : 
%The last term of the integrand in Eq.(\ref{sw2}) is $\me^{-\mi r_F^2
% \left| \mathbf{k}_\perp \right|^2/2}$. 
%Hence this integral is dominated by the region of $r_F k_\perp < 1$
% ( if $\mathbf{x}_\perp^\ms=0$ for simplicity ).
%Outside this region, the exponential rapidly oscillates and the
% integral vanishes.

\subsection{Results}

\begin{figure}
\plotone{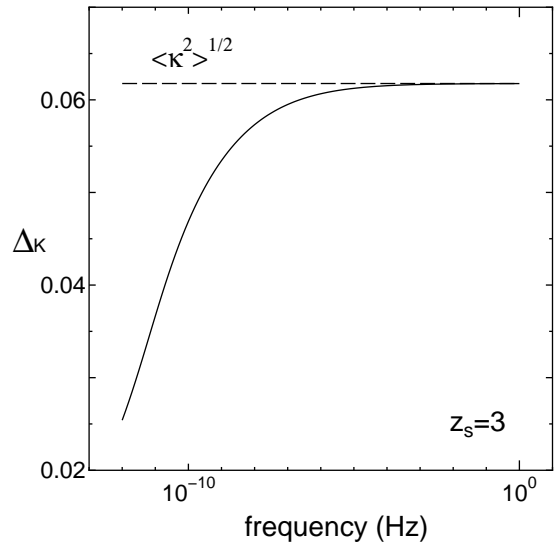}
\caption{The rms amplitude fluctuation $\Delta_K$ as a
 function of the frequency.
The source redshift is $z_\ms=3$.
The solid line is $\Delta_K$, and the dashed line is the result in the
 geometrical optics limit $\langle \kappa^2 \rangle^{1/2}$.}
\label{ca2}
\end{figure}

We show the rms (root-mean-square) amplitude fluctuation $\Delta_K$
 in Eq.(\ref{dela}) as a function of
 the frequency in Fig.\ref{ca2}. The source redshift is $z_\ms=3$.
The solid line is $\Delta_K$ and the dashed line is the geometrical 
 optics limit $\left< \kappa^2 \right>^{1/2}$.
The difference between these two lines is small for the
 LISA frequency band ($10^{-5}$ to $10^{-1}$ Hz).
But for very low frequency $f < 10^{-10}$Hz, $\Delta_K$ is clearly
 smaller than $\langle \kappa^2 \rangle^{1/2}$.
Thus the frequency band for the pulsar timing array $f \approx 10^{-9}$ Hz
 (e.g. Jenet et al. 2005), $\Delta_K$ should be used instead of
 $\langle \kappa^2 \rangle^{1/2}$. 
The wave amplitude is magnified due to lensing by the density perturbation
 on a scale larger than the Fresnel scale (see Sec.3.1). 
The integration of Eq.(\ref{dela}) is mainly contributed on the scale of
 $0.1-1$ Mpc (since $P_\delta k^2$ has a peak there).
For $f < 10^{-10}$ Hz, the Fresnel scale is $r_F > 0.4 \mbox{Mpc}
 (f/10^{-10} \mbox{Hz})^{-1/2}$ which is larger than the contributed
 scale ($0.1-1$ Mpc), and hence the amplitude is not magnified.

\begin{figure}
\plotone{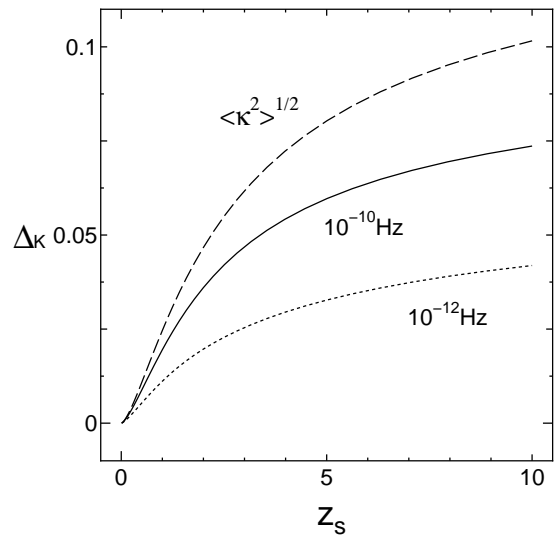}
\caption{Same as Fig.\ref{ca2}, but as a function of $z_\ms$.
The dashed line is $\langle \kappa^2 \rangle^{1/2}$, and the solid (dotted)
 line is $\Delta_K$ for the frequency of $10^{-10}$ ($10^{-12}$) Hz.}
\label{ca7}
\end{figure}

Fig.\ref{ca7} is the same as Fig.\ref{ca2}, but as a function of $z_\ms$.
The dashed line is the geometrical optics limit
 $\langle \kappa^2 \rangle^{1/2}$, the solid (dotted) line is for the 
 frequency of $10^{-10}$ ($10^{-12}$) Hz.
For $z_\ms=1-10$ the rms amplitude fluctuation is $1-10 \%$ for $f > 10^{-10}$
 Hz, while it decreases less than $5 \%$ for $f < 10^{-12}$ Hz.

\subsection{Diffraction effect in the amplitude fluctuation}

To investigate the diffraction effect in the amplitude fluctuation, 
 we expand $K(\omega)$ in terms of $1/\omega$.
From Eqs.(\ref{as}) and (\ref{sw2}), we have
\beq
 K(\omega) = 2 \omega \int_0^{r_\ms} dr \sin \left( \frac{1}{2}
 r^2_F \nabla_\perp^2 \right) U(r, \frac{r}{r_\ms} \bfx_\perp^\ms),
\eeq
where $\sin( r^2_F \nabla_\perp^2 /2 )$ $= \sum_{n=1}^{\infty} (-1)^{n+1}$
 $(r^2_F \nabla_\perp^2 /2)^{2n-1}$ $/ (2n-1)!$.
The leading term ($n=1$) is the convergence $\kappa$ in Eq.(\ref{asgeo}).
The $n$-th term ($n \geq 2$), being of the order of
 $(r_F^2 \nabla_\perp^2)^{2(n-1)} \kappa \sim (r_F / k_\perp)^{4(n-1)}
 \kappa$, is a correction term due to the diffraction effect.
The perturbation on a scale smaller than the Fresnel scale affects
the diffraction.

\section{Phase Fluctuation}

\subsection{Effects of the Phase Fluctuation in Chirp Signal}

Next, we discuss the effects of the phase fluctuation.
We consider the inspiraling BH binaries as the sources.
As the binary system loses its energy due to gravitational radiation,  
 the orbital separation decreases and the orbital frequency increases.
Hence, the frequency of the gravitational waves increases with time
 ($df/dt > 0$).
This is called a chirp signal.
We consider the frequency to be swept from $f_1$ to $f_2$.
For the binary masses $M_1$ and $M_2$ at redshift $z_\ms$, 
 the frequency of 1yr before the final merging is
\beq
  f_1=4.1 \times 10^{-5} \left( \frac{\mathcal{M}_z}{10^6 M_\odot}
 \right)^{-5/8} \mbox{Hz}  
\eeq
where $\mathcal{M}_z=(M_1M_2)^{3/5} (M_1+M_2)^{-1/5} (1+z_\ms)$ is
 the redshifted chirp mass.
The frequency at the final merging is
\beq
 f_2=4.4 \times 10^{-3} \left( \frac{M_z}{10^6 M_\odot}
 \right)^{-1} \mbox{Hz}  
\eeq
where $M_z=(M_1+M_2) (1+z_\ms)$ is the redshifted total mass.
The difference in the phase fluctuation between $f_1$ and $f_2$ in the
 chirp signal is important.

The phase fluctuation $S$ is reduced to the time delay $\omega t_d$ in
 the geometrical optics limit from Eq.(\ref{asgeo}).
But, we note that the time delay is physically unimportant, since  
 it means an arrival time shift and hence it does not change the
 waveform.
Hence we use $S-\omega t_d$ instead of $S$ as the phase fluctuation.
This quantity $S-\omega t_d$ has larger value for smaller frequency.
We define $\Delta_S$ as the rms phase difference between the
 two frequencies, $\omega_1$ and $\omega_2$ :
\beq
 \Delta_S^2(\omega_1, \omega_2) \equiv \langle \left[ \left( S(\omega_1)-
 \omega_1 t_d \right)
 - \left( S(\omega_2)- \omega_2 t_d \right)
 \right]^2 \rangle.
\label{dels}
\eeq
This is the same as $\Delta_K^2$ in Eq.(\ref{dela}) but the
 filter function is replaced with 
\beq
  F_S= \left[ \frac{\cos \left( r_{F1}^2 k_\perp^2 /2 \right) -1}
 {r_{F1}^2 k_\perp^2 /2}
% \right.  \nonumber \\ \left.
 - \frac{\cos \left( r_{F2}^2 k_\perp^2 /2 \right) -1}
 {r_{F2}^2 k_\perp^2 /2} \right]^2,
\label{fs}
\eeq
where $r_{F1}$ and $r_{F2}$ are the Fresnel scales in Eq.(\ref{fres}) for
 $\omega_1$ and $\omega_2$, respectively.

\begin{figure}
\plotone{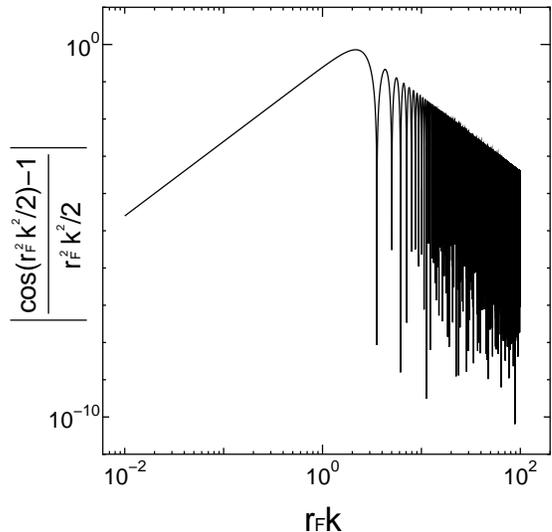}
\caption{$\left| \left[ \cos ( r_F^2 k^2 /2 )-1 \right] /
 ( r_F^2 k^2 /2 ) \right|$ as a function of $r_F k$.}
\label{fsfig}
\end{figure}

We show the behavior of first and second terms in Eq.(\ref{fs}),
 $\left| \left[ \cos ( r_F^2 k^2 /2 )-1 \right] /
 ( r_F^2 k^2 /2 ) \right|$, as a function of
 $r_F k$ in Fig.\ref{fsfig}.
From this figure, the function peaks at $r_F k \sim 1$.
Hence the phase fluctuation $S-\omega t_d$ is affected by the density
 perturbation of the Fresnel scale. 
In the chirp signal, as the frequency increases, the GW feels the
 perturbation of the smaller scale.

\subsection{Results}

In table \ref{table1}, we show the rms phase differences $\Delta_S$ in
 Eq.(\ref{dels}) for the LISA
 frequency band ($10^{-5}$ to $10^{-1}$ Hz) with $z_\ms=1,3$ and $10$.
We consider the frequency to be swept from $f_1$ to $f_2$ in the chirp signal. 
The values are in units of radian.
This table shows that the typical values of $\Delta_S$ are
 $\approx 10^{-3}$ radian. 
The results weakly depend on $f_2$ if $f_1 \ll f_2$.
This is because $S-\omega t_d$ in Eq.(\ref{dels}) has larger (smaller)
 value for smaller (higher) frequency.

\begin{deluxetable}{cccc}  
\tablecaption{rms phase differences in chirp signal $f_1 \to f_2$ 
 Hz with $z_\ms=1,3,10$. The unit is radian.}
\tablewidth{0pt}
\tablehead{
$f_1 \to f_2$ (Hz) & $z_\ms=1$ &  $3$  &  $10$
}
\startdata
 $10^{-5} \to 10^{-4}$ & $1.4 \times 10^{-3}$ &  $3.7 \times 10^{-3}$
  &  $6.5 \times 10^{-3}$ \\
 $10^{-5} \to 10^{-3}$ & $1.5 \times 10^{-3}$ &  $3.9 \times 10^{-3}$ 
  &  $6.9 \times 10^{-3}$ \\
 $10^{-4} \to 10^{-3}$ & $9.5 \times 10^{-4}$ &  $2.6 \times 10^{-3}$
  &  $4.4 \times 10^{-3}$  \\
 $10^{-4} \to 10^{-2}$ & $1.0 \times 10^{-3}$ &  $2.7 \times 10^{-3}$
  &  $4.7 \times 10^{-3}$  \\
 $10^{-3} \to 10^{-2}$ & $6.4 \times 10^{-4}$ &  $1.7 \times 10^{-3}$ 
  &  $3.0 \times 10^{-3}$  \\
 $10^{-3} \to 10^{-1}$ & $6.7 \times 10^{-4}$ &  $1.8 \times 10^{-3}$ 
  &  $3.1 \times 10^{-3}$  
\enddata
%\tablecomments{}
\label{table1}
\end{deluxetable}

\begin{figure}
\plotone{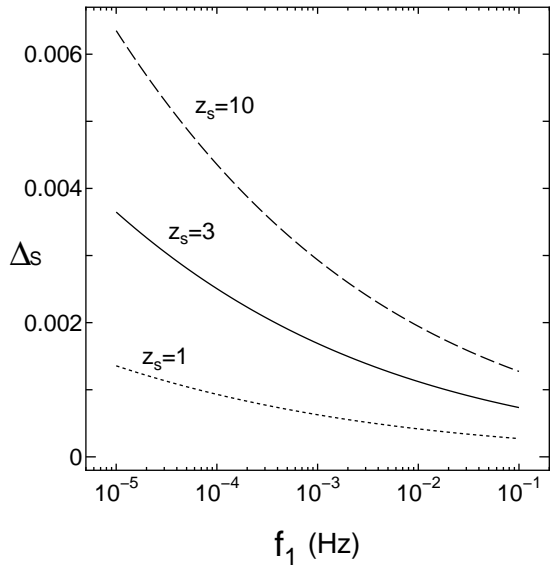}
\caption{$\Delta_S$ as a function of $f_1$ in the limit of $f_2 \to \infty$.
The source redshifts are $z_\ms=1$ (dotted line), $3$ (solid) and 
 $10$ (dashed).}
\label{cgp4}
\end{figure}

In Fig.\ref{cgp4}, $\Delta_S$ is shown as a function of $f_1$ in the limit
 of $f_2 \to \infty$.
The source redshifts are $z_\ms=1$ (dotted line), $3$ (solid) and 
 $10$ (dashed).
In order to study the behavior of $\Delta_S$, we assume a single power law
 for the power spectrum, $P(r,k) \propto k^{n}$.
The index is $n \sim -2.7$ at $k^{-1} \sim 100$ pc.
Inserting this $P(r,k)$ into Eqs.(\ref{dela}) and (\ref{fs}), we have
 $\Delta_S \propto \omega_1^{(n+2)/4} \propto \omega_1^{-0.18}$. 
With this argument and the results in table \ref{table1},
 $\Delta_S$ in $f_1 \ll f_2$ is roughly fitted by
\beq
 \Delta_S \approx 3 \times 10^{-3} ~\mbox{rad} \left( \frac{f_1}
 {10^{-4} \mbox{Hz}} \right)^{-0.18} ~\mbox{for}~ z_\ms=3.
\eeq
The above value, $3 \times 10^{-3}$ rad, is replaced by
 $1(5)\times10^{-3}$ rad for $z_\ms=1(10)$.

\subsection{Implications for GW Observations}

In the matched filtering analysis, the phase of waveform can be measured
 to an accuracy approximately equal to the inverse of the signal-to-noise
 ratio, $(S/N)^{-1}$.
Here, $S/N$ is typically $\approx 10^3$ for the SMBH mergers detected by LISA.
But including the effect of the phase fluctuation, the phase cannot be
 determined with an accuracy of less than $\Delta_S$.
Hence if $S/N$ is larger than $\Delta_S^{-1} = 10^3 (\Delta_S/10^{-3}
 \mbox{rad})$,
 the phase fluctuation becomes important and the phase of the waveform
 can be determined with an accuracy of $\sim \Delta_S^{-1}$.

\subsection{Diffraction effect in the phase fluctuation}

Similar to Sec. 3.3, we rewrite the phase fluctuation as
\beq
 S(\omega) = -2 \omega \int_0^{r_\ms} dr \cos \left( \frac{1}{2}
 r_F^2 \nabla^2_\perp \right) U(r,\frac{r}{r_\ms} \bfx^\ms_\perp).
\eeq
In $\omega \to \infty$, the above result is $\omega t_d$ in
 Eq.(\ref{as}). 
To study the behavior of the correction terms, we consider
 a simple model.
We assume the density fluctuation is confined in a thin lens plane
 at distance $r_\ell$ and is fitted by a single power law near the
 Fresnel scale : $\tilde{U}(r,\bfk) \propto k_\perp^\alpha \delta(k_r)
 \delta(r-r_\ell)$ where $\alpha$ is an index. 
Then, from Eqs.(\ref{as}) and (\ref{sw2}), we have
\beq
 S(\omega) = \omega t_d + C \omega^{\alpha/2+2}.
\eeq
The coefficient $C$ is constant and the second term is of the order
 of $10^{-3}$ rad at the LISA frequency band.
Hence, by measuring the phase fluctuation directly, one would obtain
 the constant $C$ and the index $\alpha$.
% the density perturbation of the Fresnel scale.

\if{}
For example, the phase in the Newtonian quadrupole formula is
 $\Psi(f)=2 \pi f t_\mc$ 
 $- \phi_\mc - \pi/4$ $+3/4 (8 \pi \mathcal{M}_z f)^{-5/3}$ where
 $t_\mc$ and $\phi_\mc$ are coalescence time and phase.
From the above argument, we have $\Delta \Psi \approx \mbox{max}[(S/N)^{-1},
 \Delta_\ms]$.
Hence, the relative error for the chirp mass is
\beqa
 \frac{\Delta \mathcal{M}_z}{\mathcal{M}_z} \approx 5.3 \times 10^{-4}
 \left( \frac{\mathcal{M}_z}{10^6 M_\odot} \right)^{5/3}
 \left( \frac{f}{10^{-4} \mbox{Hz}} \right)^{5/3} \nonumber \\
 \times \mbox{max} \left[ (S/N)^{-1}, \Delta_\ms \right]
\eeqa
\fi{}

\section{Validity of the Born Approximation}

Throughout this paper we assume the weak density fluctuation and
 employ the Born approximation to discuss the lensing effects on
 the waveform.
Since the variances of the amplitude and
 the phase fluctuations are much smaller than one,
 this approximation is valid in almost all cases. 
But in a few cases, the GW may pass through the strong density fluctuation
 or pass near massive compact objects (i.e. strong lensing).
Hence we have some comments about the validity of the Born approximation.

For the amplitude fluctuation $K$, the maximum of $K$ is the convergence
 $\kappa$ from Fig.\ref{ca2}.
For the thin lens plane at $r_\ell$, $\kappa$ is the surface density
 of the lens $\Sigma$ divided by the critical density $\Sigma_{\mbox{cr}}
 = (1/4 \pi) r_\ms / r_\ell r_{\ell \ms} \simeq
 2 \times 10^3 M_\odot {\mbox{pc}}^{-2} [(r_\ms / r_\ell r_{\ell \ms}) / 
 {\mbox{Gpc}}^{-1}]$ (here $r_{\ell \ms}=r_\ms-r_\ell$).
Hence if the GWs pass through high density region $\Sigma >
 \Sigma_{\mbox{cr}}$ such as a core of galaxy or cluster, the Born
 approximation breaks down and one should use the Kirchhoff diffraction
 integral to obtain the exact waveform (Schneider, Ehlers \&
 Falco 1992, Sec.4.7 and 7). 

%Similarly, the phase fluctuation is $S-\omega t_d = r_F^2 \nabla^2_\perp 
% \kappa /4 + \mathcal{O}(r_F^4 \nabla^4_\perp  \kappa)$ for the thin lens.  
%Since $\nabla^2_\perp \kappa \sim \kappa/L^2$ ($L$ is the lens size), 
%  the high density $\Sigma > \Sigma_{\mbox{cr}}$ with the compact 
% size $L < r_F$.

If the gravitational potential is a Gaussian random field, one can
 obtain exact solutions of the correlation functions of the lensed
 waveform (Macquart 2004; see also Ishimaru 1978, Ch.20).
If there are many compact lens objects and the GWs are scattered several 
 times, the exact lensed waveform can be obtained by using the
 multiple lens-plane theory in the wave optics (Yamamoto 2003).

\section{Conclusion \& Discussion}

We have discussed the lensing effects on the amplitude and phase
 of a waveform.
The rms amplitude fluctuation is $1-10 \%$ for $f > 10^{-10}$ Hz, which is
 the same as the result in weak lensing.
But for a very low frequency of $f < 10^{-12}$ Hz, it decreases
 to less than $5 \%$.
In the chirp signal, the phase fluctuation is typically $10^{-3}$ radian
 at the LISA frequency band.
The phase cannot be measured with an accuracy less than this value.

The power spectrum $P_\delta$ has been measured down to a small scale
 $k^{-1} \sim \mbox{several} \times 0.1~\mbox{Mpc}$ from the Ly$\alpha$
 forest (e.g. Zaroubi \etal 2005).
In this paper, we assume the formula of $P_\delta$ is valid down to
 the Fresnel scale $k^{-1} \sim 100$ pc.
If the amplitude or the phase fluctuation is measured in future,
 the constraints for $P_\delta$ at $\sim 100$ pc could be obtained.

\acknowledgments

I would like to thank Naoshi Sugiyama and Naoki Seto for useful comments
 and discussion.
I also thank anonymous referee for useful comments and references
 about electromagnetic scattering.

\appendix

\section{Conformal transformation}

The relation between the Einstein tensor $G_{\mu \nu}$ for the metric
 $g_{\mu \nu}$ and $\tilde{G}_{\mu \nu}$ for $\tilde{g}_{\mu \nu} (= a^2 
 g_{\mu \nu})$ is given by \cite{w84},
\beqa
 \tilde{G}_{\mu \nu} = G_{\mu \nu} -2 \nabla_\mu \nabla_\nu \ln a
 + 2 \left( \nabla_\mu \ln a \right) \left( \nabla_\nu \ln a \right)
 \nonumber \\
 + g_{\mu \nu} \left( \nabla^\sigma \ln a \right)
 \left( \nabla_\sigma \ln a \right) + 2 g_{\mu \nu} \nabla^\sigma
 \nabla_\sigma \ln a.
\label{a1}
\eeqa
Let us consider the linear perturbation $\tilde{h}_{\mu \nu}
 (=a^2 h_{\mu \nu})$ in the
 background metric $\tilde{g}^B_{\mu \nu} (=a^2 g^B_{\mu \nu})$
 given in Eq.(\ref{metric}):
\beq
 \tilde{g}_{\mu \nu} = \tilde{g}^B_{\mu \nu} + \tilde{h}_{\mu \nu}
 = a^2 \left( g^B_{\mu \nu} + h_{\mu \nu} \right).
\label{a2}
\eeq
Inserting Eq.(\ref{a2}) into (\ref{a1}), we obtain the linearizing
 Einstein equation
\beq
 \delta \tilde{G}_{\mu \nu} = \delta G_{\mu \nu} + 2 \frac{a^{\prime}}{a}
 \delta \Gamma^0_{\mu \nu} + \left[ 2 \frac{a^{\prime \prime}}{a}
 - \frac{a^{\prime 2}}{a^2} \right] \delta \left( g_{\mu \nu}
 g^{0 0} \right) -2 \frac{a^{\prime}}{a} \delta \left( g_{\mu \nu}
 g^{\sigma \rho} \Gamma^0_{\sigma \rho} \right),
\eeq
where $a^{\prime}=da/d\eta$, $\Gamma$ is the Christoffel symbol, and
 $\delta$ means the perturbed component. 
The first term $\delta G_{\mu \nu}$ is of the order of
 $|h_{\mu \nu}|/\lambda^2$ ( $\lambda$ is the wavelength of the gravitational
 wave ). 
The other terms on the right-hand side are roughly
 $|h_{\mu \nu}|/(\lambda \lambda_H)$ or $|h_{\mu \nu}|/\lambda_H^2$ where
 $\lambda_H$ is the horizon scale, since $a^{\prime} = a^2 H \sim
 a^2/\lambda_H$. 
Hence, if $\lambda \ll \lambda_H$ the propagation equation for the
 gravitational wave is conformally invariant, i.e. $\delta
 \tilde{G}_{\mu \nu} = \delta G_{\mu \nu} =0$.

\end{document}